\documentclass[nofootinbib,twocolumn,showpacs,preprintnumbers,pre,aps]{revtex4-1}

\usepackage{graphicx}
\usepackage{bm}
\usepackage{amsmath}
\usepackage{amssymb}

\begin{document}

\title{Budding of domains in mixed bilayer membranes}%

\author{Jean Wolff$^{1,3}$}\email{jean.wolff@ics-cnrs.unistra.fr}

\author{Shigeyuki Komura$^{2}$}\email{komura@tmu.ac.jp}

\author{David Andelman$^{3}$}\email{andelman@post.tau.ac.il}

\affiliation{$^{1}$Institut Charles Sadron, UPR22-CNRS 23, \\
rue du Loess BP 84047, 67034 Strasbourg Cedex, France\\
$^{2}$Department of Chemistry, Graduate School of Science and Engineering,\\
Tokyo Metropolitan University, Tokyo 192-0397, Japan\\
$^{3}$Raymond and Beverly Sackler School of Physics and Astronomy,\\
Tel Aviv University, Ramat Aviv, Tel Aviv 69978, Israel}

\date{December 26, 2014}

\begin{abstract}
We propose a model that accounts for budding behavior of domains in lipid
bilayers, where each of the bilayer leaflets has a coupling between its local
curvature and local lipid composition.
The compositional asymmetry between the two monolayers leads to an overall
spontaneous curvature.
The membrane free-energy contains three contributions: bending energy, line tension,
and a Landau free-energy for a lateral phase separation.
Within a mean-field treatment, we obtain various phase diagrams which
contain fully-budded, dimpled and flat states.
In particular, for some range of membrane parameters,
the  phase diagrams exhibit a tricritical behavior as well as three-phase
coexistence region.
The global phase diagrams can be divided into three types and are analyzed in
terms of the curvature-composition coupling parameter and domain size.
\end{abstract}

\maketitle

\section{Introduction}
\label{sec:introduction}

Biological membranes are multi-component assemblies  typically composed of
lipids, cholesterol, glyco-sugars, and proteins, whose presence is indispensable to the
normal functioning of living cells~\cite{AlbertsBook}.
Given the complexity of biological membranes, studies of model membranes
have been conducted {\it in vitro} in order to gain insight on the structural and physical
behavior of biomembranes. Many studies, in particular over the last two decades,
have focused on simplified artificial systems containing vesicles in solution, composed of
ternary mixtures of lipids and cholesterol~\cite{VK05,SK_DA_Review}.
By decreasing temperature, the ternary mixtures undergo a
phase separation between a liquid-ordered (L$_{\rm o}$) phase
and a liquid-disordered (L$_{\rm d}$) one~\cite{VK,YIMKO}.
Depending on thermodynamical parameters, the liquid domains show
one of three distinct domain shapes: \textit{flat},
\textit{dimpled} (partially budded), or \textit{fully-budded}~\cite{UrsellBook}.

A theoretical model for domain-induced budding of planar membranes
was proposed by Lipowsky~\cite{Lipowsky92,Lipowsky93}, and later
was extended for the case of closed vesicles~\cite{JL93,JL96}.
In the model, the competition between  membrane bending-energy and
domain line-tension leads to a budding transition under the constraint
of fixed domain area. 
Hu \textit{et al}.~\cite{hu} proposed a mechanism based on this interplay, 
which stabilizes patterns of several domains on closed vesicles without requiring
any osmotically induced membrane tension.

Returning to the case of planar membranes, Lipowsky's model~\cite{Lipowsky14} predicts that:
(i) an initially flat domain deforms spontaneously into a completely spherical
bud when the initial domain size exceeds a critical size; and (ii) dimpled domains
are stable only when the spontaneous curvature of the bilayer membrane is nonzero.
The latter prediction was later re-examined~\cite{UKP}, because dimpled domains
are observed experimentally in vesicles with no apparent spontaneous curvature~\cite{baumgart}.
In order to resolve this discrepancy, Rim \textit{et al.}~\cite{RUPK}
considered the effect of adding an overall  lateral tension acting on the membrane,
and used ideas about entropy-driven lateral tension that were originally
proposed by Helfrich and Servuss~\cite{HS}.
The resulting phase diagram~\cite{RUPK} contains regions
of stability for the three different domain morphologies as mentioned above.
Moreover, the effect of the lateral tension on  budding  was discussed by Lipowsky 
and coworkers~\cite{Lipowsky92, lischi}. 
In particular, the budding process requires that the activation energy has to exceed 
the energy barrier associated with  the surface tension.

In this paper, we propose a model that describes domain-induced budding in
bilayers composed of a binary mixture of lipids.
We suggest that dimpled domains can be formed and remain stable due to a possible
asymmetry between the two monolayer compositions.
We show that the dimpled structure appears when the line tension along the domain rim
is not too large.
Global phase diagrams are calculated within mean-field theory and, in some range
of system parameters, we obtain a tricritical behavior as well as three-phase
coexistence region.
We discuss different morphologies that characterize the phase diagrams in
terms of  model parameters.

It has been recognized long time ago that such an asymmetry in monolayer composition
leads to a nonzero bilayer spontaneous curvature~\cite{SPA,SPAM}.
The coupling between composition and monolayer curvature was also
considered~\cite{MS} in order to describe the transition between lamellar and
vesicular phases of bilayer membranes composed of two types of amphiphiles.
It is worthwhile mentioning  related works by Harden \textit{et al.}~\cite{HM94,harden} and
G\'o\'zd\'z \textit{et al.}~\cite{GG01}, who studied budding and domain shape
transformations in bilayer membranes.
In Refs.~\cite{HM94,harden}, the phase separation is assumed to occur only in
one of the monolayers, and the domain spontaneous curvature due to the compositional
asymmetry is kept constant.
For finite spontaneous curvatures, it was shown that the dimpled domains are
obtained in equilibrium when both line and surface tensions are small~\cite{harden}.
More recently, ring-shaped domains were experimentally obtained in model membranes
by using a bud-mimicking topography~\cite{Ryu}.
Such a ring-shaped domain is located around a bud-neck region having a negative
curvature, and is characterized by the composition asymmetry between the two
monolayers.

\begin{figure}[tbh]
\begin{center}
\includegraphics[scale=0.4]{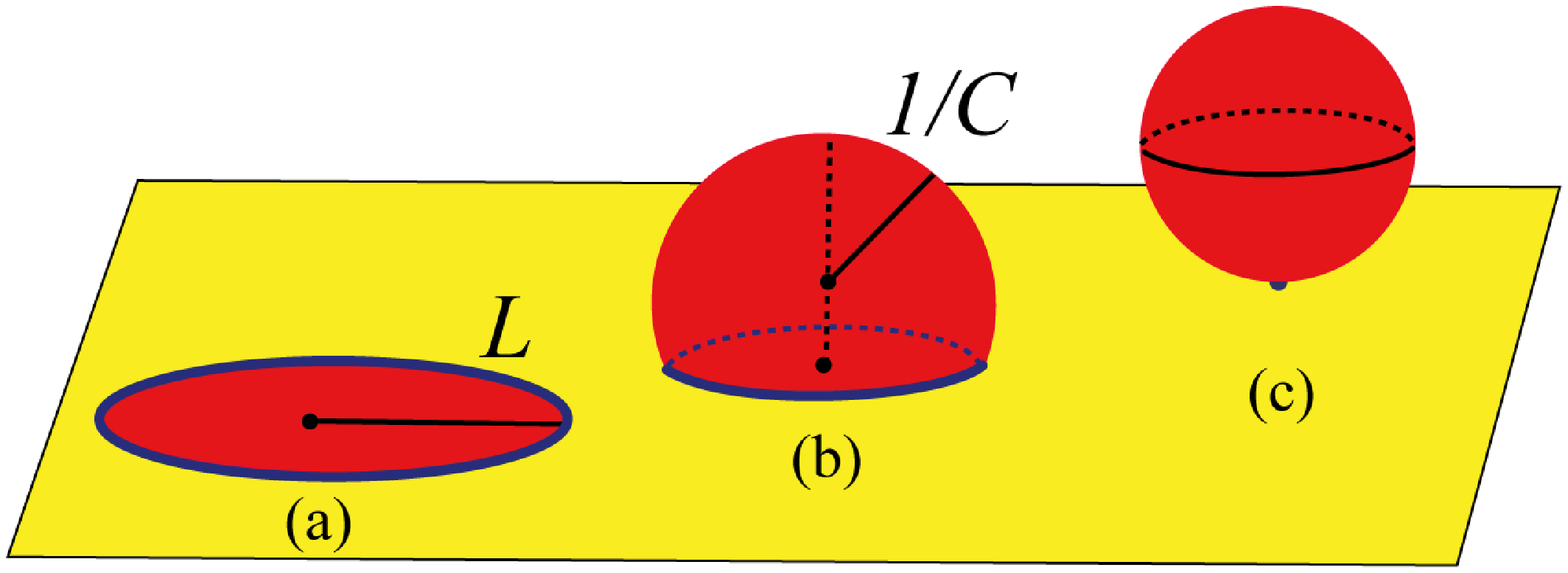}
\end{center}
\caption{\textsf{
(a) The flat phase, (b) the dimpled phase,  and (c) the fully-budded phase.
In (a) the circular flat domain (red)
has a radius $L$ and area $S=\pi L^2$.
In (b) a bud of the same area $S$ forms a spherical cap of radius $1/C$, where
$C$ is the curvature, embedded in an otherwise flat membrane.
In (c) a fully-budded domain of area $S$ has a spherical shape, just touching the flat membrane.
The line tension $\gamma$ acts along the boundary
(blue line) between the domain and the flat membrane.
}}
\label{fig1}
\end{figure}

The outline of our paper is as follow.
In the next section, we present a model for bilayer domains.
In Sec.~\ref{sec:diagram}, various mean-field phase diagrams are obtained by
changing the ratio between the domain size and the invagination length,
as well as tuning the inter-monolayer coupling parameter strength.
Finally, Sec.~\ref{sec:discussion} includes some discussion and interpretation of our
results.

\begin{figure}[tbh]
\begin{center}
\includegraphics[scale=0.7]{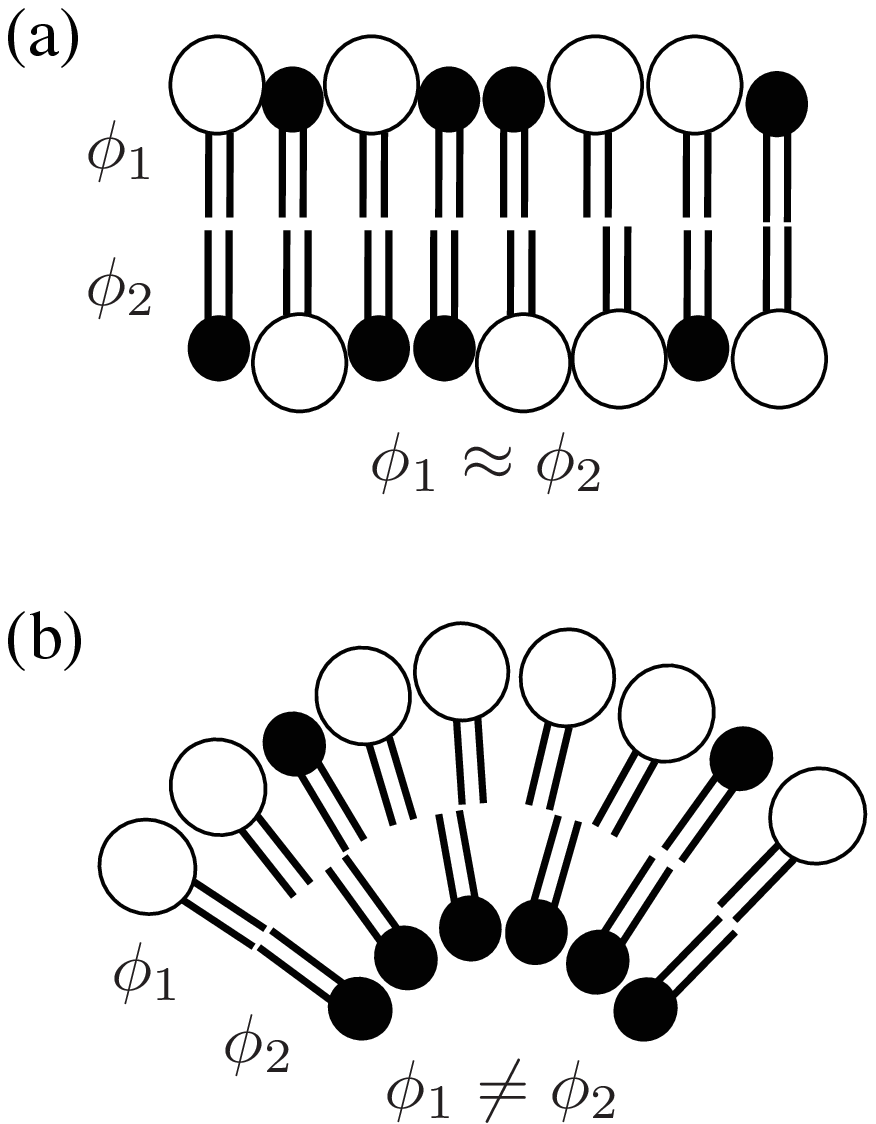}
\end{center}
\caption{\textsf{(a) Flat bilayer domain when the relative A/B compositions in the two monolayers,
$\phi_1$ and $\phi_2$, are almost symmetric, $\phi_1 \approx \phi_2$.
(b) Curved bilayer domain when the compositions are asymmetric,
$\phi_1 \neq \phi_2$.
The spontaneous curvature of each monolayer is assumed to depend linearly on the composition,
as given by Eq.~(\ref{spontaneous}).
}}
\label{fig2}
\end{figure}

\section{Model}
\label{sec:model}

We model the membrane as a bilayer having two monolayers (leaflets),
each composed of an A/B
mixture of lipids that can partition themselves asymmetrically between the
two monolayers.
We consider the case where the lipids can undergo a lateral phase separation
creating domains rich in one of the two components.
As discussed below, these domains can also deform (bud) in the normal direction, and
the deformations are controlled by the membrane curvature elasticity.
Because we do not include any gradient terms
in the free energy, it results in an unrealistic discontinuous jump
of the membrane curvature close to the bud edge.
For the fully-budded state this jump in curvature does not matter because
the bud neck corresponds to a small length scale of the order of the membrane thickness.
However, in the dimpled state, this jump occurs on a bigger length scales and
artificially affects the free energy.
We will further introduce a coupling between local lipid composition and local
curvature~\cite{SPA,SPAM,MS}, which can eventually drive the budding process of the
membrane.

We start by considering a single two-dimensional (2D) circular domain of an initial and arbitrary
radius $L$ embedded in an otherwise flat (2D) membrane, as shown in Fig.~\ref{fig1}(a).
The area of the domain,  $S=\pi L^2$, is assumed to stay constant even when the
domain buds into the third dimension.
For simplicity, we consider only budded domains whose shape is a spherical cap of
radius $1/C$ (Fig.~\ref{fig1}(b)).
The total bending energy of the budded domain is given by adding the curvature
contributions from the two monolayers~\cite{Helfrich73,SafranBook}:
\begin{equation}
E_{\rm b} =2 \pi L^2 \kappa \left[ (C -C_0)^2
+ (C + C_0)^2 \right],
\label{bend}
\end{equation}
where $\kappa$ is the bending rigidity modulus and $C_0$ the monolayer spontaneous
curvature.
As shown in Fig.~\ref{fig2}, the two monolayers are fully coupled together, and their
curvatures are given by $+C$ and $-C$, respectively.

The next contribution is the domain edge energy that is proportional to the edge length and
its line tension, $\gamma$~\cite{Lipowsky92}:
\begin{equation}
E_{\rm ed} = 2 \pi L \gamma\sqrt{1-(LC/2)^2}.
\label{edge}
\end{equation}
In the extreme case, when the domain buds into a complete spherical domain as in
Fig.~\ref{fig1}(c), $C=\pm 2/L$ and $E_{\rm ed}=0$.

For domains that are composed of two different lipid types, the relative
composition in each monolayer is defined as $\phi_i=\phi_i^{\rm A}-\phi_i^{\rm B}$
($i=1,2$), where $\phi_i^{\rm A}$ ($\phi_i^{\rm B}$) is the molar fraction of the
${\rm A}$ lipid ($\rm B$ lipid) in the $i$-th monolayer.
We assume that each of the monolayers is incompressible, hence, $\phi_i^{\rm A}$+$\phi_i^{\rm B}=1$.
For simplicity sake, the molecular areas of A and B species are taken to be the same, meaning that the
molar fraction of the lipids is indenting to their area fraction.
As in any A/B mixture, the possibility of a phase separation due to partial
incompatibility between the two species can be described by a phenomenological
Landau expansion of the free energy in powers of $\phi_i$ around the critical point, $\phi_i=0$.
In our case, this expansion is done separately for
each monolayer, and the free energy is the sum of the two contributions:
\begin{equation}
E_{\rm ph} = \pi L^2 \frac{U}{\Xi^2}
\sum_{i=1}^2 \left[
\frac{t}{2} \phi_i^2 + \frac{1}{4} \phi_i^4 - \mu \phi_i
\right],
\label{phase}
\end{equation}
where $\Xi \equiv \kappa/\gamma$ is the invagination length,
$U$ a parameter that sets the energy scale,
$t \sim (T-T_{\rm c})/T_{\rm c}$  the reduced temperature ($T_{\rm c}$ being
the critical temperature), and $\mu$ the chemical potential that fixes
the A/B relative composition in each layer.
In general, a different chemical potential can be assigned to each of the two monolayers.
However, since it is difficult to control the average composition in each layer
separately, we introduce only one chemical potential $\mu$ that fixes the
total relative composition $\phi_1+\phi_2$ of the entire bilayer.
Notice that we allow exchange of lipid molecules between the two
monolayers via a flip-flop process.
Bilayers where each of the monolayer compositions can be controlled independently
will be addressed in our future work.

As argued before~\cite{MS}, we do not include any $\phi_i$ gradient term in
Eq.~(\ref{phase}) because we consider only homogeneous composition within a single domain.
The energy cost associated with a gradient term in composition is effectively taken into account
through the line tension $\gamma$ in Eq.~(\ref{edge}), which is regarded here as an
external control parameter. 
This assumption of $\gamma$ can be justified for situations of strong segregation (far 
from the critical point) between the domain and the background, for which the domain 
boundary is sharp.

Hereafter, we will use several dimensionless variables: a rescaled curvature $c \equiv LC$,
rescaled spontaneous curvature $c_0 \equiv LC_0$, and rescaled invagination length
$\xi\equiv \Xi/L$.
The coupling between the spontaneous curvature $c_0$ and composition is taken into
account by assuming a linear dependence on $\phi_i$~\cite{MS} (see also Fig.~\ref{fig2}):
\begin{equation}
c_0(\phi_i)=\bar{c}_0-\beta \phi_i,
\label{spontaneous}
\end{equation}
where all variables in Eq.~(\ref{spontaneous}) are dimensionless,
$\bar{c}_0$ is the spontaneous curvature of the monolayer at its
critical composition $\phi_i=0$, and $\beta$ a coupling constant.
Since $\bar{c}_0$ is a constant that merely shifts the origin of the chemical potential
$\mu$, we can drop it without loss of generality.

The total free-energy of the bilayer model is given by the sum of Eqs.~(\ref{bend}), (\ref{edge}),
and (\ref{phase}):
\begin{equation}
E_{\rm tot}=E_{\rm b}+E_{\rm ed}+E_{\rm ph}.
\end{equation}
Denoting the average and difference of the two monolayer
compositions, respectively, by
\begin{equation}
\phi_{+}\equiv \frac{\phi_2+\phi_1}{2}~~~{\rm and}~~~
\phi_{-}\equiv \frac{\phi_2-\phi_1}{2},
\end{equation}
the dimensionless total free-energy of one domain, $\varepsilon=E_{\rm tot}/2\pi \kappa$,
is expressed as
\begin{align}
\varepsilon & =
2 c^2 - 4 \beta c \phi_{-} + 2 \beta^2 (\phi_{+}^2+\phi_{-}^2)
\nonumber \\
& + \frac{1}{\xi} \sqrt{1-c^2/4}
\nonumber \\
& + \frac{1}{\xi^2} \left( \frac{U}{2 \kappa} \right)
\biggl[
t(\phi_{+}^2+\phi_{-}^2)
\nonumber \\
& +\frac{1}{2} (\phi_{+}^4+6\phi_{+}^2 \phi_{-}^2+\phi_{-}^4) -2\mu \phi_{+}
\biggr],
\label{total}
\end{align}
where we have dropped unimportant constant terms.
Within a mean-field theory, the equilibrium state of the system and the phase transitions
are determined by minimization of the above $\varepsilon$ with respect
to $\phi_\pm$ and $c$.

We note that Eq.~(\ref{total}) depends on three dimensionless parameters: $\beta$,
$\xi$, and $U/2 \kappa$, while
the thermodynamic variables are the temperature $t$, and the three
order parameters: $\phi_\pm$ and $c$.
In the calculations presented hereafter, we set $U/(2 \kappa)=1$ and vary the values of $\beta$ and $\xi$.
Since the total free energy is invariant under simultaneous exchange of
$\beta \rightarrow -\beta$ and $\phi_{-} \rightarrow -\phi_{-}$, it is sufficient
to study only the $\beta > 0$ range.

Typical experimental values of flat domain size are in the range of
$L\simeq 50$ -- $500$\,nm~\cite{SI},
the bending rigidity $\kappa \simeq 10^{-19}$\,J $\approx 25 k_{\rm_B}T$~\cite{SafranBook},
and  line tension in the range of
$\gamma \simeq 0.2$ -- $6.2 \times 10^{-12}$\,J/m~\cite{baumgart,tian}.
These parameter values yield invagination length, $\Xi$, of the order
$0.01L$ to $10L$ ($\xi \simeq 0.01$ -- $10$), as will be used in the next section.

\section{Phase Behavior and Phase Diagrams}
\label{sec:diagram}

\begin{figure}[tbh]
\begin{center}
\includegraphics[scale=0.45]{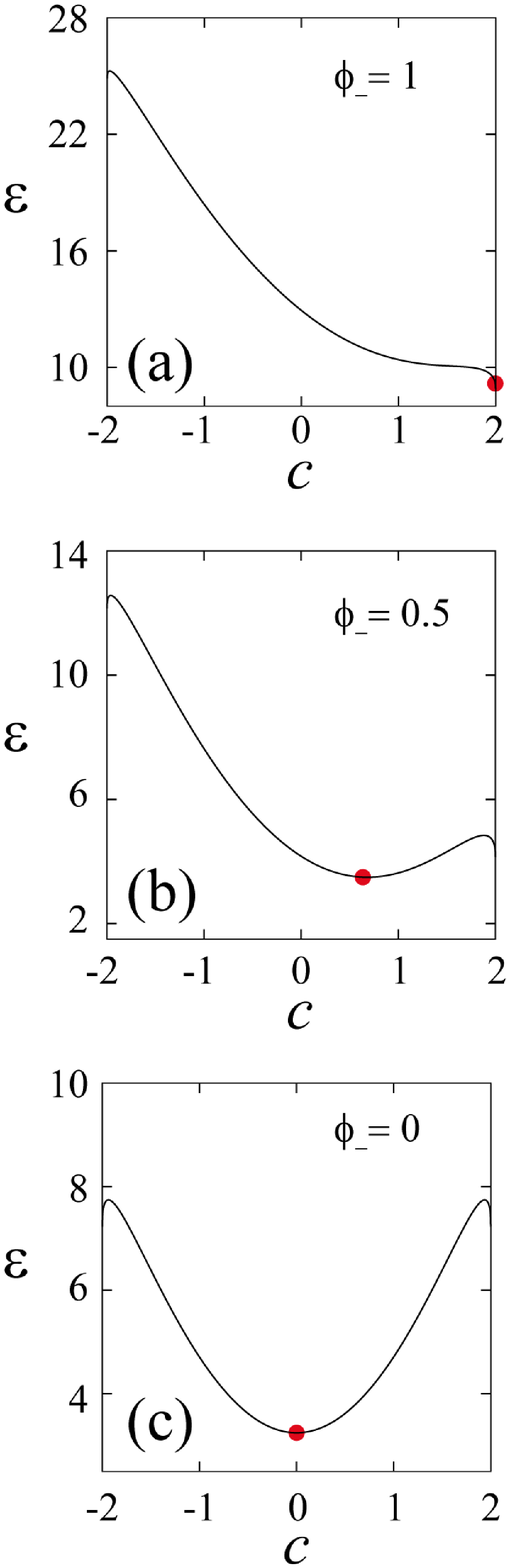}
\end{center}
\caption{\textsf{Plots of the free energy $\varepsilon$ as a function of the curvature
$c$ for (a) $\phi_-=1$, (b) $\phi_-=0.5$, and (c) $\phi_-=0$.
The other parameter values are: $t=-0.5$, $\phi_{+}=0.4$, $\beta=1$, and $\xi=0.25$.
The free-energy minimum is shown by a red dot, and corresponds to
the fully-budded state ($c=2$), dimpled state ($0<c<2$), and flat state
($c=0$),  in (a), (b) and (c), respectively.
}}
\label{fig3}
\end{figure}

\begin{figure}[tbh]
\begin{center}
\includegraphics[scale=0.36]{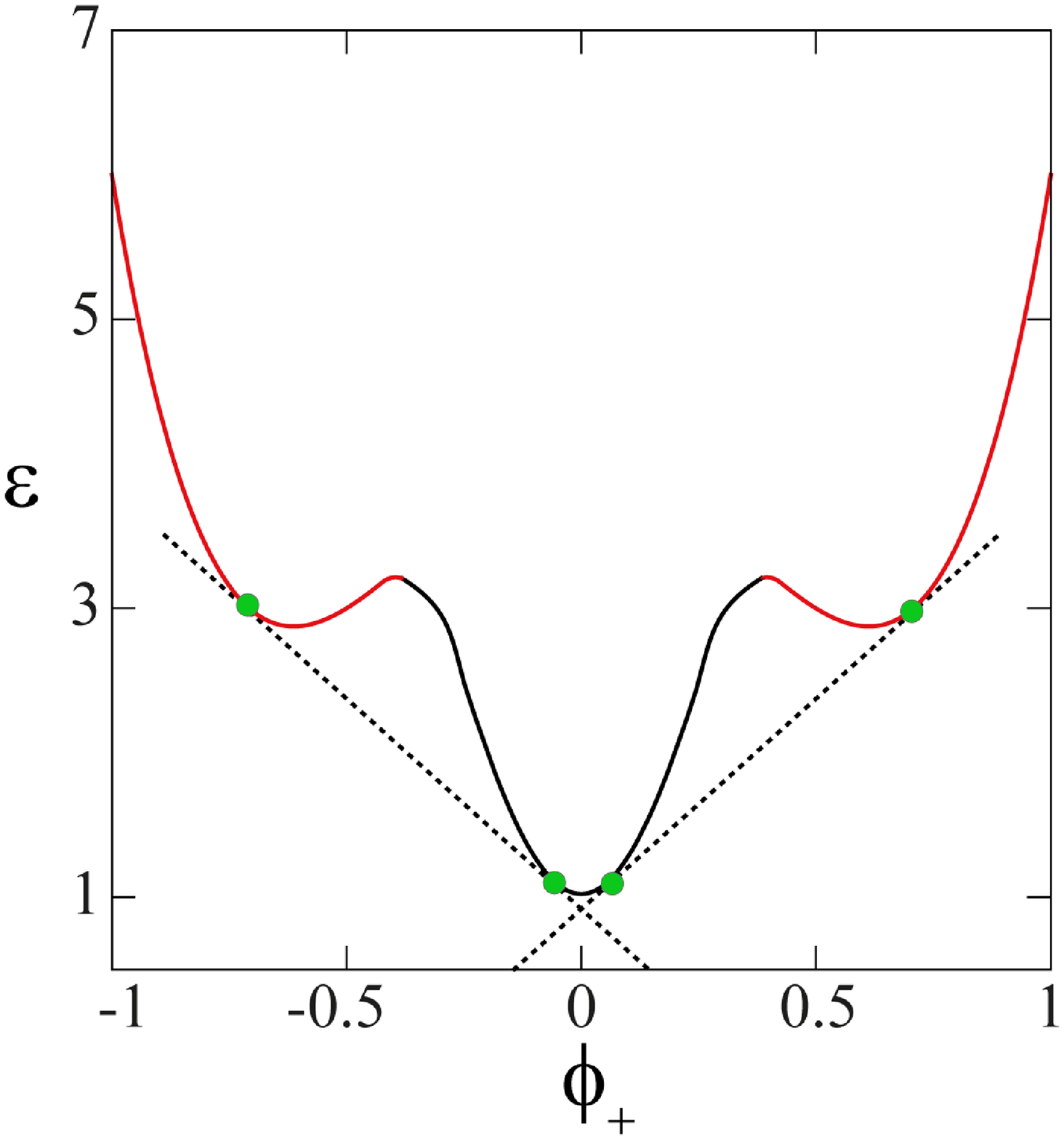}
\end{center}
\caption{\textsf{Plot of the free energy $\varepsilon$ as a function of $\phi_{+}$ for
$t=-0.5$, $\beta=1$, and $\xi=0.25$.
The red and black lines of the free energy correspond to the flat and fully-budded states,
respectively.
The two dashes lines are the common tangent constructions,
which determine two sets of coexisting
compositions indicated by green dots.
}}
\label{fig4}
\end{figure}

\subsection{Flat, dimpled and fully-budded states}
\label{sec:define}

The total free energy $\varepsilon$ in Eq.~(\ref{total}) is first minimized
with respect to the curvature $c$, yielding
\begin{equation}
4c - 4 \beta\phi_{-} - \frac{c}{4\xi\sqrt{1-c^2/4}}=0.
\label{minimizec}
\end{equation}
The above equation indicates that the value of $c$, taken at the minimum of
$\varepsilon$, uniquely determines the value of $\phi_{-}$, as long  as $|c|<2$.
The value of the curvature determines which of the domain states is the equilibrium one:
flat (F) with $c=0$, fully-budded (B) with $c=\pm 2$,
or dimpled (D) with $0< \vert c \vert <2$.
By substituting back the above minimization condition for $\vert c \vert <2$
into the total free energy, we obtain $\varepsilon(\phi_{+}, \phi_{-})$ as a function
of $\phi_{+}$ and $\phi_{-}$.
This free energy is further minimized with respect to $\phi_{-}$, leading to an
expression $\varepsilon(\phi_{+})$ that is only a function of $\phi_{+}$.
We assume that the average composition $\phi_{+}$ is a conserved order-parameter
(while $c$ and $\phi_{-}$ are non-conserved), and can be controlled by varying the
conjugate chemical potential $\mu$ acting as a Lagrange multiplier.

In order to illustrate this minimization process,  we plot
$\varepsilon$ as a function of the curvature
$c$ in Fig.~\ref{fig3}, for given values of $t$ and $\phi_{\pm}$.
We see that the free energy takes its minimum at different curvature values
(marked by red circles) for different $\phi_-$ values.
Figure~\ref{fig3}(a), (b) and (c) correspond to the fully-budded, dimpled and flat
states, respectively.
In Figure~\ref{fig4} the free energy $\varepsilon$ that was minimized with respect to both $c$ and
$\phi_-$ is plotted as a function of $\phi_+$ for a fixed temperature.
Different colors of the free energy plot correspond to different domain states
(F and B).
The two dashes lines are the common tangents that determine the two sets
of coexisting compositions.
For the chosen parameter values as in Fig.~\ref{fig4},
the flat and fully-budded phases are in coexistence (F+B and B+F).

\subsection{Phase diagrams}
\label{sec:phase}

\begin{figure}[tbh]
\begin{center}
\includegraphics[scale=0.34]{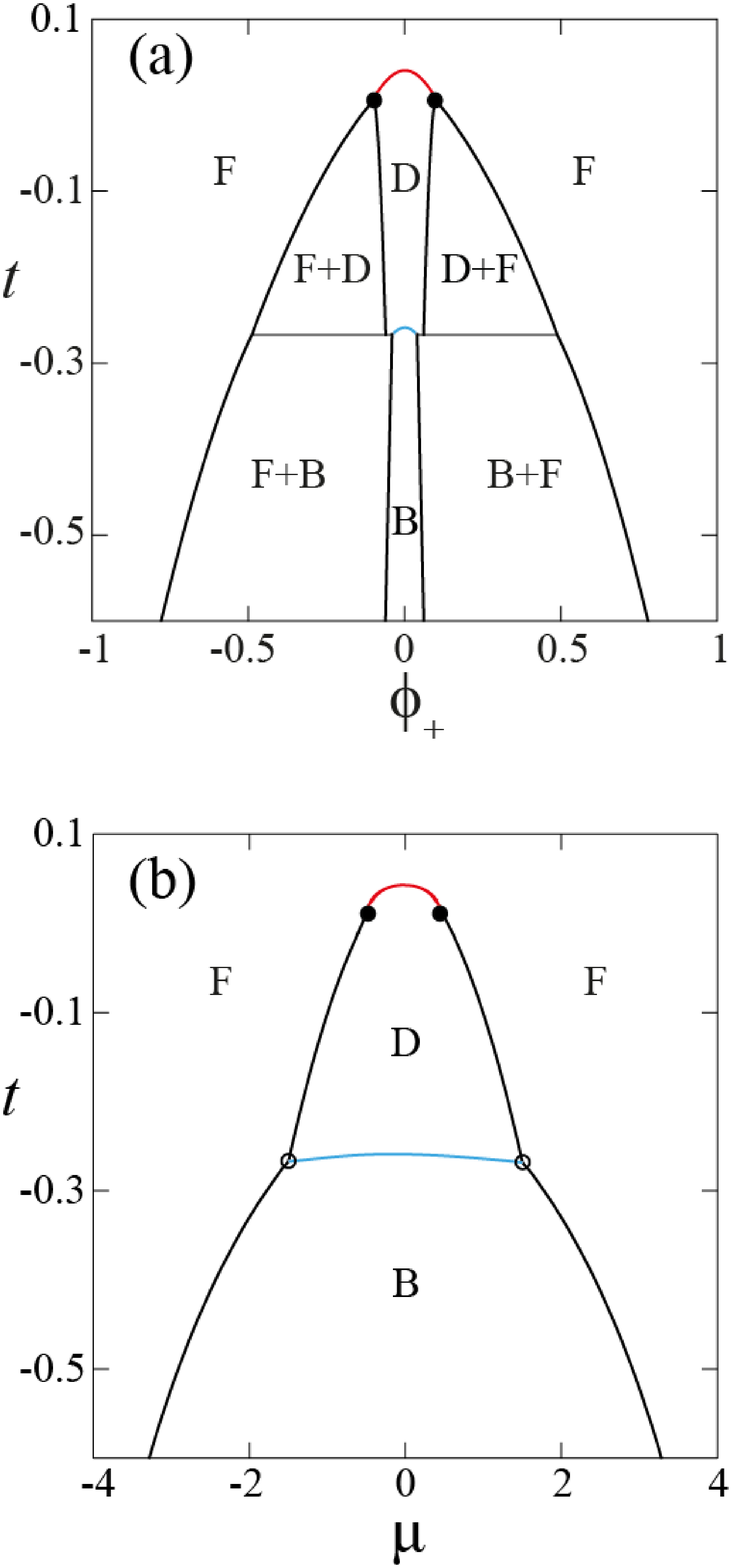}
\end{center}
\caption{\textsf{(a) Phase diagram in the ($\phi_{+}$, $t$) plane, where $\phi_{+}$ is the average
composition and $t$ the reduced temperature; and, (b) in the ($\mu$, $t$) plane, where
$\mu$ is the chemical potential.
The parameters are $\beta=1$ and $\xi=0.25$.
``F", ``D", and ``B" stand, respectively, for flat, dimpled, and fully-budded phases.
Coexistence regions are denoted by ``F+D" \textit{etc.} in (a).
The black and red lines indicate first- and second-order phase transitions,
respectively, while the blue line indicates  a first-order phase transition
with a discontinuous jump in both $c$ and $\phi_{-}$.
The filled circles represent the tricritical points
($t_{\rm tcp}\simeq 0.011$,
$\phi_{+}^{\rm tcp}\simeq \pm 0.094$,
$\mu_{\rm tcp}\simeq \pm 0.488$),
and the open circles in (b) represent the triple points
($t_{\rm tri}\simeq -0.267$,
$\mu_{\rm tri}\simeq \pm 1.49$),
where three phases coexist, with $\phi_{+}^{\rm tri}=\pm 0.484, \pm 0.066$,
and $\pm 0.045$.
}}
\label{fig5}
\end{figure}

\begin{figure}[tbh]
\begin{center}
\includegraphics[scale=0.43]{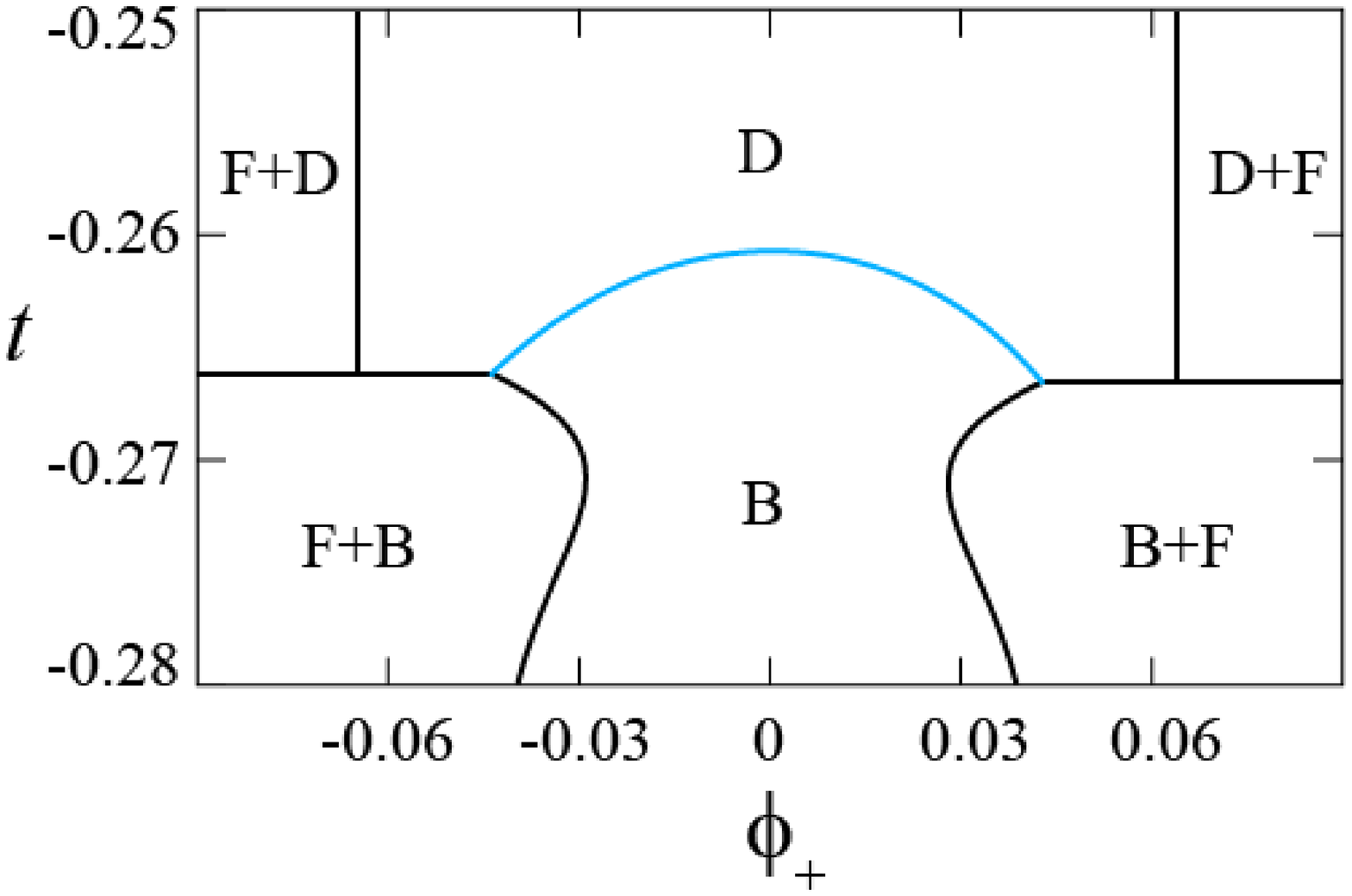}
\end{center}
\caption{\textsf{Enlarged middle-zone ($\phi_{+}\approx 0$ and $t$ around the triple point value)
of Fig.~\ref{fig5}(a).
The blue line is a first-order phase transition between dimpled and fully-budded phases,
with a jump in the curvature value from $c\simeq 0.75$ to $c=2$.
The solid black lines indicate the boundaries of the two-phase coexistence regions:
F+D and F+B.
On the triple line, $t_{\rm tri}\simeq -0.267$, three phases coexist with
$\phi_{+}^{\rm tri}=\pm 0.066$ and $\pm 0.045$.
}}
\label{fig6}
\end{figure}

The numerically-computed phase diagrams are three-dimensional ones for fixed values of
$\xi$ and $\beta$. They can be plotted either in the $(\phi_{+},t,c)$ or  $(\mu,t,c)$ parameter space.
We recall that the throughout
this work we set for simplicity, $U/(2\kappa)=1$.
In addition, note that the equilibrium $\phi_-$ value is self-determined by the
equilibrium $c$ value according to Eq.~(\ref{minimizec}).
As it is too cumbersome to present 3D plots, we plot 2D phase diagrams in the
$(\phi_{+}, t)$ or $(\mu, t)$ planes, which represent a projection in the $c$ direction,
or 2D cuts in the $(c,t)$ plane for fixed values of the conserved order-parameter,
$\phi_{+}$ (see Fig.~\ref{fig8}).

In Fig.~\ref{fig5} we present the phase diagrams that are obtained numerically
for $\beta=1$ and $\xi=0.25$.
In (a) the phase diagram is plotted in the $(\phi_{+},t)$ plane, and in (b) in
the $(\mu,t)$ plane.
The phase diagram in (a) is symmetric about $\phi_{+}=0$, and in (b)
about $\mu=0$.
At high temperatures only the flat phase is stable.
For lower temperatures, in the range $-0.267 <t< 0.042 $, the dimpled phase becomes
stable.
The phase diagrams show a tricritical behavior, similar to the well-known tricritical behavior
of Blume--Emery--Griffiths spin--one model~\cite{BEG}.

The red line in Fig.~\ref{fig5} denotes a second-order phase transition between F and D
phases, occurring when $c \rightarrow 0$.
It terminates at two symmetric tricritical points (filled circles),
$t_{\rm tcp}\simeq 0.011$, $\phi_{+}^{\rm tcp}\simeq \pm 0.094$ in (a), and
$\mu_{\rm tcp}\simeq \pm 0.488$ in (b).
The tricritical points are also obtained analytically using some approximations and their
calculated values, $t_{\rm tcp}\simeq 0.014$ and $\phi_{+}^{\rm tcp}\simeq \pm 0.095$,
agree well with the numerical ones.
More details on the analytical derivations are provided in the Appendix.
For $t<t_{\rm tcp}$, the phase transition between F and D becomes first-order (solid
black line) with coexistence lines in the $(\mu,t)$ plane and two coexistence regions,
marked as F+D and D+F in the $(\phi_{+},t)$ plane.
As one crosses this phase transition line, there is a jump in $\phi_{+}$, as well as in
$c$ and $\phi_{-}$, and the jump in $\phi_{-}$ is fully determined by a similar jump in $c$.

Two triple points are shown as open circles in Fig.~\ref{fig5}(b) at
$t_{\rm tri}\simeq -0.267$ and $\mu_{\rm tri}\simeq \pm 1.49$, or equivalently as
a horizontal line in Fig.~\ref{fig5}(a).
At the triple point, the three phases (F, D and B) coexist.
In order to explain in more detail the phase behavior close to the triple line, we show
in Fig.~\ref{fig6} an enlarged section of Fig.~\ref{fig5}(a) around the triple line.
The tip of the middle (blue) line starts at about $t\simeq -0.261$ and terminates
at the triple-point temperature.
This is a first-order phase transition line where both $c$ and $\phi_-$ have a
discontinuous jump from their dimpled values  ($c\simeq 0.75$) to their fully-budded values ($c=2$).
The other solid lines delimit the two-phase coexistence regions: F+D above the triple
line and F$+$B below it.

\begin{figure}[tbh]
\begin{center}
\includegraphics[scale=0.32]{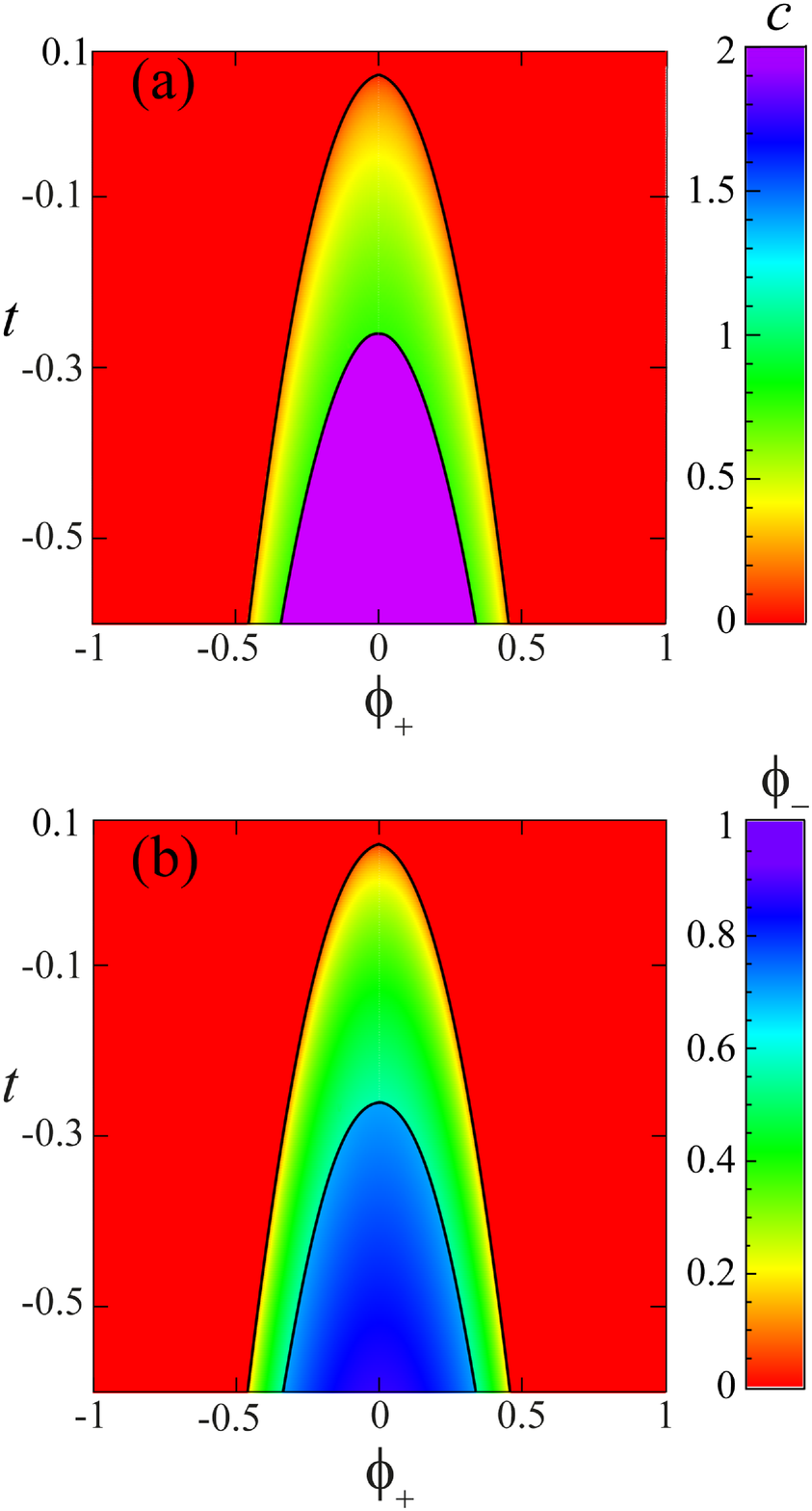}
\end{center}
\caption{\textsf{(a) The curvature $c$ as a contour plot in the ($\phi_{+}$, $t$) plane,
with a color bar that corresponds to $0\le c \le 2$. A jump from $c=2$ to about
$0.75$ can be seen as $t$ increases its value and eventually crosses the lower black line, while along
the upper black line $c$ vanishes continuously.
In (b) the compositional asymmetry between the two monolayers, $\phi_{-}$,
is plotted as a contour plot in the ($\phi_{+}$, $t$) plane.
As $t$ increases, a jump is seen from $\phi_-=0.7$ to $0.55$ along the lower black curve.
For the upper black curve $\phi_-$  vanishes continuously (just as $c$ was).
The chosen parameters are $\beta=1$ and $\xi=0.25$.
}}
\label{fig7}
\end{figure}

\begin{figure}[tbh]
\begin{center}
\includegraphics[scale=0.4]{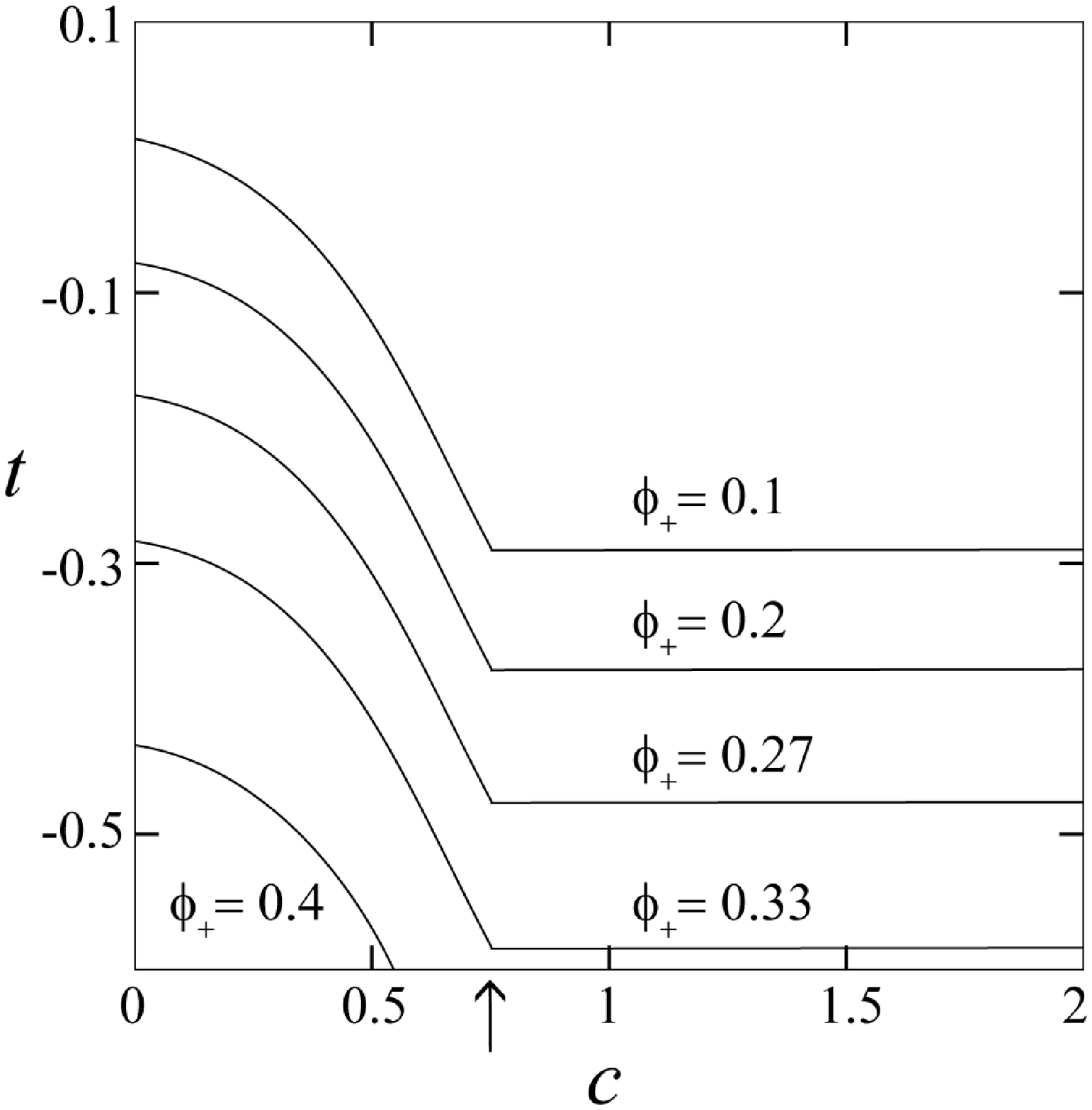}
\end{center}
\caption{\textsf{The reduced temperature $t$ as function of
the equilibrium curvature $c$ for fixed values of
$\beta=1$, $\xi=0.25$, and several values of $\phi_{+}=0.1, 0.2, 0.27, 0.33$ and $0.4$.
As $t$ decreases, the curvature $c$ first continuously increases from
zero, and then discontinuously jumps from $c \simeq 0.75$ (marked by an arrow
on the $c$-axis) to $c=2$.
The former corresponds to the second-order phase transition from the flat state to the
dimpled one, while the latter to the first-order phase transition from the
dimpled state to the fully-budded one.
}}
\label{fig8}
\end{figure}

In Fig.~\ref{fig7}(a), we plot the equilibrium values of $c$, and
those of $\phi_-$ in (b), in order to view more clearly the phase transitions.
Both $c$ and $\phi_{-}$ are plotted in the $(\phi_{+}, t)$ plane as a contour color plot.
In (a) we see two parabola-like lines delimiting different values of $c$.
At the upper black line, the curvature continuously tends towards zero, $c\rightarrow 0$.
The region close to the curve tip ($\phi_{+}\approx 0$) coincides with the second-order phase transition
between F and D phases (the red line of Fig.~\ref{fig5}(a)), while the rest of the line
lies inside the two-phase coexistence region, and does not influence the equilibrium state of
the system.
The lower blue line represents a jump in $c$ from $c \simeq 0.75$ (D phase) to $c=2$
(B phase).
Its top region (close to $\phi_+=0$) coincides with the first-order phase transition between
D and B phases (the blue lines in Figs.~\ref{fig5}(a) and \ref{fig6}), and the rest of the
line lies within the F+B coexistence region.
In Fig.~\ref{fig7}(b), a similar contour plot is shown for $\phi_-$, as is
determined by Eq.~(\ref{minimizec}).

The complementary plot is shown in Fig.~\ref{fig8} in the $(c,t)$ plane for several
fixed values of $\phi_{+}$ ranging from 0.1 to 0.4.
We recall that
$\phi_{+}=(\phi_1+\phi_2)/2=(\phi^{\rm A}_1+\phi^{\rm A}_2)/2-(\phi^{\rm B}_1+\phi^{\rm B}_2)/2$
is a conserved quantity determined by the total amount of the A and B lipids in
the domain. In the model we control it by the chemical potential $\mu$.
As  $t$ is lowered, the minimized curvature $c$
continuously increases from zero.
This represent a continuous (second-order) phase transition from the flat state (F with $c=0$) to the dimpled one (D with $c>0$).
When the temperature is lowered  even further, the curvature discontinuously jumps from
$c \simeq 0.75$ (indicated by an arrow on Fig.~\ref{fig8}) to $c=2$.
This is a first-order phase transition from the dimpled state (D) to the fully-budded (B) one.
Notice that the maximum curvature $c \simeq 0.75$ of the dimpled state does not depend
on the average composition $\phi_+$.

\begin{figure}[tbh]
\begin{center}
\includegraphics[scale=0.34]{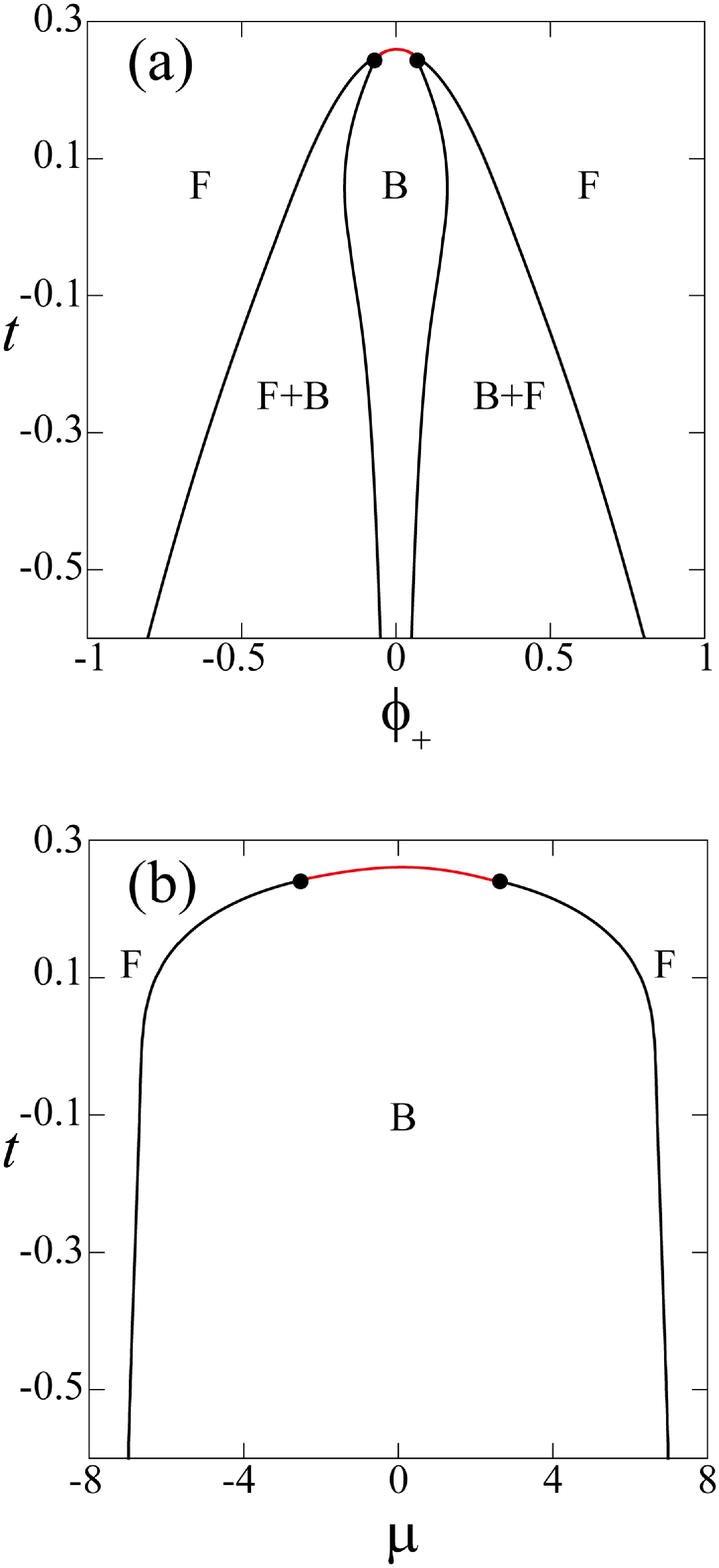}
\end{center}
\caption{\textsf{(a) Phase diagram in the ($\phi_{+}$, $t$) plane, and (b) in the ($\mu$, $t$)
plane for $\beta=1$ and $\xi=1/7\simeq 0.143$.
The meaning of the lines and symbols is the same as in Fig.~\ref{fig5}.
A critical (red) line separates the F and B phases and terminates at two tricritical points (filled circles)
with $(t_{\rm tcp}\simeq 0.240, \phi_{+}^{\rm tcp}\simeq \pm  0.063, \mu_{\rm tcp}\simeq \pm 2.64)$.
For $t<t_{\rm tcp}$, coexistence regions, B+F and F+B, separate the F and B phases.
}}
\label{fig9}
\end{figure}

When the $\xi$  value is decreased, while keeping $\beta$ fixed, the D phase disappears,
and the only remaining stable phases are F and B, with a phase transition
between them.
This is shown on Fig.~\ref{fig9} where the chosen parameter values are $\beta=1$ and $\xi=1/7\simeq 0.143$.
A second-order phase transition (red line) is seen between the F and B phases in the
proximity of the symmetric $\phi_{+}=0$ axis.
This second-order line ends at two tricritical points located at $t_{\rm tcp}\simeq 0.240$ and
$\phi_{+}^{\rm tcp}\simeq \pm 0.063$ in (a), or equivalently, $\mu_{\rm tcp}\simeq \pm 2.64$
in (b).
Below the tricritical temperature, the coexistence region is between the F and
B phases (F+B), and is delimited by the solid black lines.
Note that as the D phase disappeared there is no three-phase coexistence at these parameters values.
The disappearance of the D phase can be understood in the following way.
Smaller values of the invagination length, $\xi= \kappa/(L \gamma)$, correspond to larger
values of the line tension $\gamma$, and domains will fully bud for lower temperatures
without showing any D state.

\begin{figure}[tbh]
\begin{center}
\includegraphics[scale=0.34]{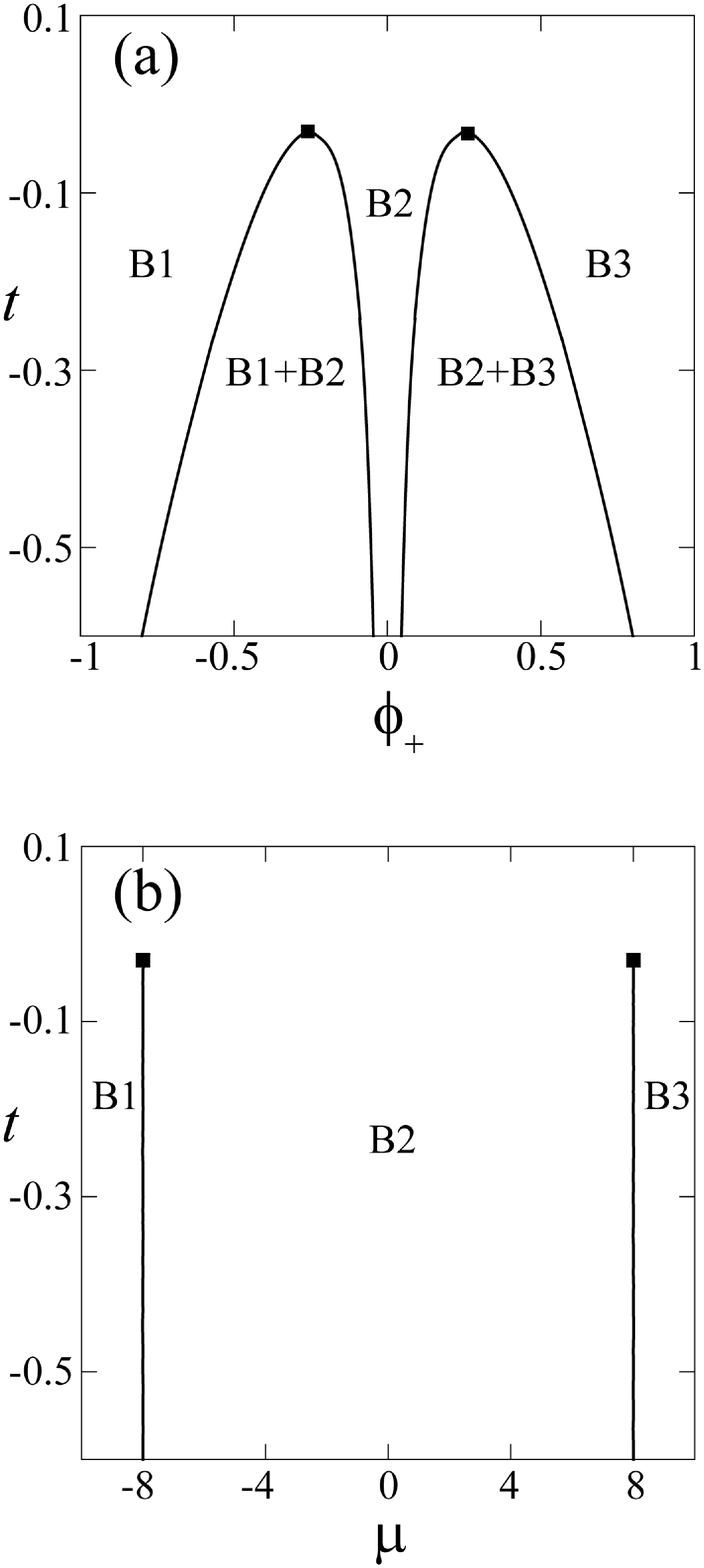}
\end{center}
\caption{\textsf{(a) Phase diagram in the ($\phi_{+}$, $t$) plane, and (b) in the ($\mu$, $t$)
plane for $\beta=1$ and $\xi=0.125$.
The meaning of the lines and symbols is the same as in Fig.~\ref{fig5}.
Only the B domain is stable and a first-order phase transition separates between
B1 and B2, and another one between B2 and B3. Each of the coexistence regions terminates at a critical
point $(t_{\rm c}\simeq -0.028, \phi_{+}^{\rm c}\simeq \pm 0.251, \mu_{\rm c} \simeq \pm 8)$.
The filled squares correspond to the critical points.
}}
\label{fig10}
\end{figure}

At yet lower values of $\xi$, the line tension is large enough  so
that only the B phase exists, while the F phase disappears.
In Fig.~\ref{fig10}, we present such a phase diagram for $\beta=1$ and $\xi=0.125$.
The only coexistence regions are between different fully-budded phases, denoted as
B1+B2 and B2+B3.
Each of these coexistence regions terminates at critical points (filled squares),
$t_{\rm c}\simeq -0.028$, $\phi_+^{\rm c}\simeq \pm 0.251$ and $\mu_{\rm c} \simeq \pm 8$.

\begin{figure}[tbh]
\begin{center}
\includegraphics[scale=0.4]{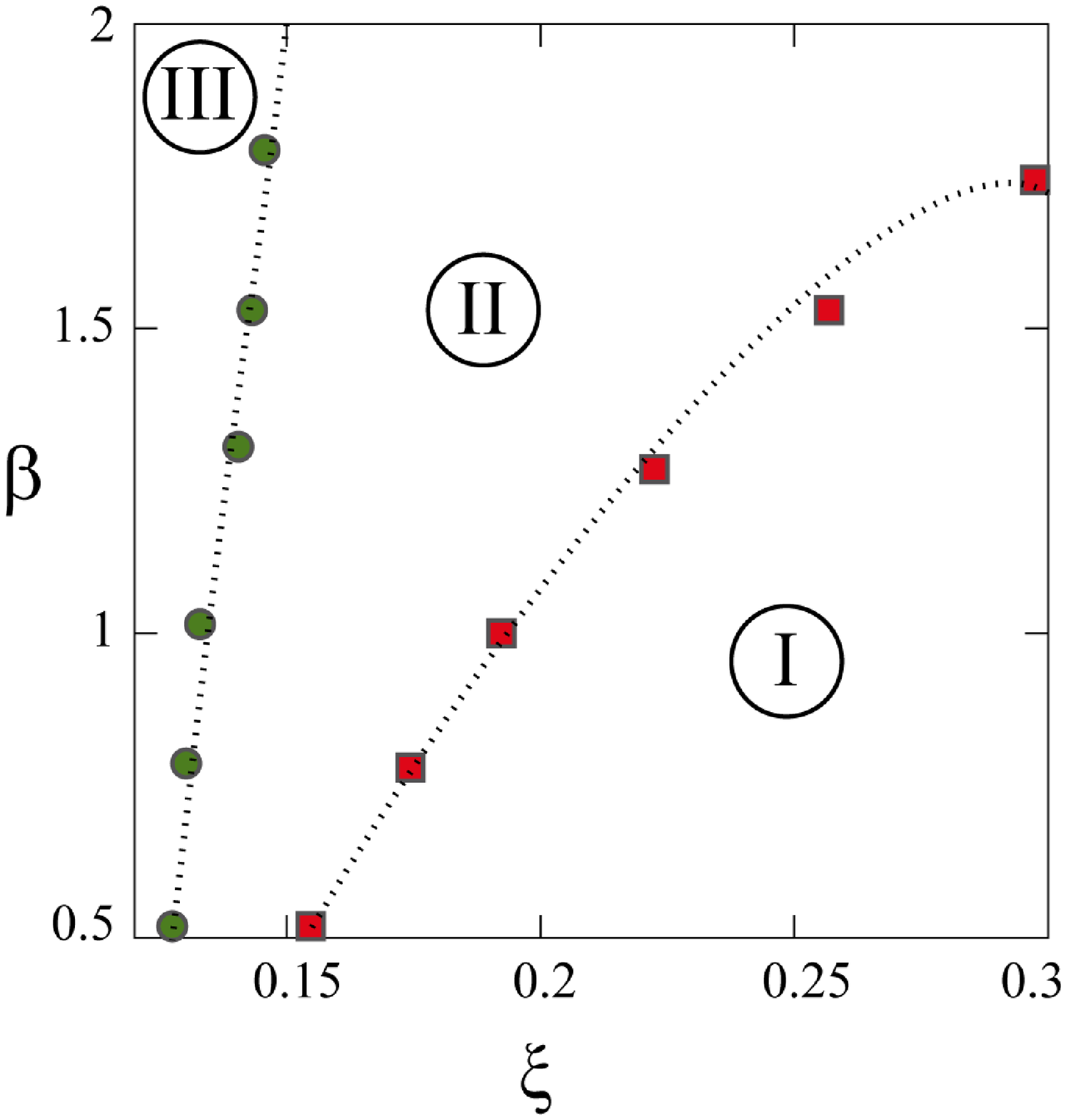}
\end{center}
\caption{\textsf{Behavior diagram as a function of $\xi=\Xi/L$ and
the monolayer coupling parameter $\beta$.
Examples of type I, II, III phase-diagram morphologies are given in
Figs.~\ref{fig5}, \ref{fig9}, and \ref{fig10}, respectively.
The red squares delimit the calculated borderline between type I and II behavior,
while the green circles delimit the calculated crossover between type II and III.
The dashed lines serve only as a guideline to the eye
and the error bar of the data points is about $\pm 0.1$ in $\beta$.
}}
\label{fig11}
\end{figure}

\subsection{Effects of $\xi$ and $\beta$ on the phase behavior}
\label{sec:global}

By exploring the entire parameter range of $\xi$ and $\beta$, we find
the crossover between the three types of phase diagrams as represented
in Figs.~\ref{fig5} (type I), \ref{fig9} (type II), and \ref{fig10}
(type III).
This is shown in Fig.~\ref{fig11}, where we present the stability regions
for each of these three phase behaviors in the ($\xi$, $\beta$) plane.
Type I is characterized by the existence of a dimpled phase, and has a
triple point where the D, B and F phases coexist.
In type II, the tricritical points exist but the triple points and the
D phase disappear.
Type III is dominated by various B phases, with coexistence regions between
them that terminate at a critical point.
The crossover line between type II and III behaviors is almost a straight line, while
the  crossover line from type II to I is almost linear  for $\beta<1.25$,
and then saturates at about $\beta\simeq 1.8$.
This saturation occurs when the coupling is strong (large $\beta$) and/or the domain
size is small (large $\xi$).
At these values, budding is promoted because of the large
spontaneous curvature.

When $\xi$ decreases, for a fixed value of the coupling parameter $\beta$,
the B phase swells and the D phase disappears, signaling
the crossover between type I and II.
Upon further decrease of $\xi$, only the B phase stays, i.e., crossover between type
II and III.
On the other hand, the larger $\beta$ is, the larger is the spontaneous curvature
that favors the fully-budded state.
For this reason, at higher values of $\beta$, the system buds at lower temperatures
for the same value of $\xi$.

\section{Discussion}
\label{sec:discussion}

We have proposed a model that accounts for domain budding of lipid bilayers,
where each of the bilayer leaflets has a coupling between its local curvature and local
A/B lipid composition.
The composition asymmetry between the two leaflets is equivalent to the introduction
of a membrane spontaneous curvature. This spontaneous curvature
is not taken to be fix (as was assumed in previous works),
but is calculated and depends on the asymmetry in leaflet composition. Hence,
due to this extra mechanism of generating a spontaneous curvature,
dimpled domains can be stabilized even for bilayers with a nominal zero spontaneous curvature.
Our free-energy model contains three contributions: bending energy accounting for domain
deformation in the normal direction, line tension along the rim of the budded or
flat domain, and a Landau free-energy expansion that accounts for a lateral phase separation
of the binary lipid mixture.
We assume that the domain area remains constant during the budding process.

Our model predicts three states for domain as were observed experimentally:
fully-budded (B), dimpled (D) and flat (F) states.
In particular, in some range of parameters, the D state is found to be the most
stable one.
The obtained results indicate that for a certain range of temperatures, monolayer composition,
domain size and coupling between curvature and composition, a triple point can appear.
At the triple point, the  B, D and F phases coexist, each with its own composition.
Such a triple point has been reported already by Harden
\textit{et al.}~\cite{harden}.
Moreover, we also found a tricritical point that corresponds to the intersection
of a critical (second-order) line, which joins a first-order phase transition region
between F and D.
Finally, three types of phase diagram morphologies are found and analyzed in terms of
the coupling parameter $\beta$ and domain size $\xi$ (see Fig.~\ref{fig11}).

Formation of domains in membranes and their understanding remain an open and
ever-challenging problem, even after an intense research in the last two decades.
Many hypotheses have been proposed to explain the domain appearance and their possible
structure and function~\cite{VK05,SK_DA_Review,LS}.

One of the important assumptions in our model is that the domain size $L$
and area $S$  remain fixed during the budding process.
Domains of fixed size can be obtained in thermodynamical equilibrium for a binary mixed membrane that
undergoes a lateral \textit{micro-phase separation}, and forms a 2D modulated phase with
an equilibrated spatial periodicity~\cite{SA}.
This a micro-phase separation can be driven by a coupling between local lipid composition
and membrane curvature, leading to a curvature
instability~\cite{leibler,andel,kawa,taniguchi}, as was in particular discussed in
Refs.~\cite{KK,kumar}.
When the A/B average lipid composition is off-critical, circular domains rich in one
of the lipid can form spontaneously and be arranged in a hexagonal array, embedded in
a background rich in the second lipid.
These circular domains are characterized by their equilibrium fixed size, and can undergo a budding
process as explored in the present work.

Formation of finite-size domains in equilibrium can also be explained by the presence
of hybrid lipids having one saturated tail and a second unsaturated one~\cite{BPS09}.
Such hybrid lipids decrease the domain line tension~\cite{YBS,PS13,PS14} and offer 
another potential mechanism to induce micro-phase separation.
In a previous work~\cite{hirose09,hirose12}, we considered a model that includes a
coupling between a compositional scalar field and a 2D vectorial order parameter.
This coupling yields an effective 2D free energy that exhibits micro-phase separation
and resulted in a modulated phase.
A somewhat different viewpoint of membrane domains has been recently discussed by
Shlomovitz \textit{et al.}~\cite{shlomovitz}, who investigated a general phenomenological model capable
of producing macro-phase separation, micro-phase separation, and
microemulsion-like phases.
In these  works, the characteristic length of compositional modulations is responsible
for the origin of finite-size domains that are equilibrium structures.
These types of domains can undergo the budding transition as we have discussed in this paper.

On the other hand, when \textit{macro-phase separation} takes place in mixed
membranes, the domain size grows to macroscopic sizes as function of time,
and the assumption of fixed domain size becomes more questionable.
However, if the shape transformation of domains occurs on time scales much faster than the time required
for domain coarsening, one can still use our equilibrium argument for domain morphologies
whereby we regard the domain size $L$ and, hence, the invagination length $\xi$,
as time dependent.
In fact, the slowdown of  dimpled-domains coarsening was experimentally
observed~\cite{YIMKO} and theoretically discussed~\cite{UKP}.
According to these works, the suppression of the phase separation may be caused by
membrane-mediated elastic interactions and/or hydrodynamic interactions acting
between domains. Our model with its assumption of fixed domain size can also be applied in such situations.

A dynamical growth of the budding domain size
was proposed~\cite{Lipowsky92} to occur in two steps,
when the spontaneous curvature is not too large.
In the early stage the domains are small and the diffusion-aggregation phenomenon
induces a growing dimpled domain until its size becomes unstable.
Whereas in the later stage, the domains are large enough to fully bud into a sphere
that detaches completely from the planar membrane at the neck point.
In other words, the budded domain curves mainly during the second step.
Our results are in qualitative agreement with these predictions.
We showed in Fig.~\ref{fig11} that, for small values of $\beta$ and for small domain
sizes (large $\xi\sim 1/L$), the typical phase diagram is of type I for which the
dimpled state appears as a stable phase.
As the domain size becomes larger (smaller $\xi$), the typical phase diagram
will change to either type II or III so that the membrane can bud easily.
In contrast, for large values of $\beta$,  the dimpled state cannot be stable.
In this case, the domain will retain a highly curved state already in the early
stage of the phase separation, and the budding can occur only in one step.

Finally, our model may be applied to describe the formation and growth
of vesicles in mixed amphiphilic systems~\cite{SH}.
For example, it was observed in experiment that mixtures of anionic and cationic surfactants
in solution form disk-like bilayers for some range of relative surfactant composition.
As these disk-shaped bilayers grow in size, they transform into spherical caps and eventually
become spherically closed vesicles.
Such a sequence of morphological changes was indeed observed by cryo-TEM (transmission
electron microscopy)~\cite{SH}.
In such a setup, it is likely that the spontaneous curvature of bilayer membranes are induced
due to the compositional asymmetry between the two monolayers.
Hence, one can expect that disks, caps, and vesicles can be analyzed similarly to the flat,
dimpled, and fully-budded phases in our model.

\begin{acknowledgments}

We thank T.\ Kato, B.\ Palmieri and S.\ A.\ Safran for useful discussions and 
numerous suggestions.
JW acknowledges support from the Service de Coop\'eration Scientifique et Universitaire de
l'Ambassade de France en Isra\"el, the French O.R.T association, and O.R.T school of Strasbourg.
SK acknowledges support from  Grant-in-Aid for Scientific Research on Innovative Areas
``Fluctuation \& Structure" (grant No.\ 25103010), grant No.\ 24540439 from the MEXT of Japan,
and the JSPS Core-to-Core Program ``International research network for non-equilibrium dynamics
of soft matter".
DA acknowledges support from the Israel Science Foundation (ISF)
under grant No.\ 438/12 and the US-Israel Binational Foundation (BSF)
under grant No.\ 2012/060.
\end{acknowledgments}

\appendix*
\section{The tricritical point}
\label{tcp}

It is possible to compute analytically the location of the tricritical point
($t_{\rm {tcp}}$), corresponding to the intersection of the first- and second-order
transition lines in the phase diagram of Fig.~\ref{fig5}(a).
The left side of the binodal line corresponds to $c=0$ (F phase), while
its right side corresponds to $c > 0$ (D phase).
Using the fact that the F phase with $c=0$ has two symmetric monolayers, and that
$\phi_{-}=0$ from Eq.~(\ref{minimizec}), we calculate the free energy for $c=\phi_{-}=0$
by substituting $c=0$ in Eq.~(\ref{total}):
\begin{equation}
\varepsilon(\phi_{+}, c=0)=\frac{1}{2}\xi^{-2}\phi_{+}^4 +(2\beta^2+t\xi^{-2})\phi_{+}^2
+ {\xi}^{-1}.
\label{eq:eq31}
\end{equation}
The free-energy expression can then be expanded up to fourth order in $c$ (valid
close to the tricritical point where  $c\ll 1$), yielding
\begin{eqnarray}
\varepsilon(c,\phi_{+}, \phi_{-})&=& \frac{1}{2}\xi^{-2}(\phi_{+}^4+\phi_{-}^4+6\phi_{+}^2\phi_{-}^2)\nonumber\\
& &+ (2\beta^2+t\xi^{-2})(\phi_{+}^2+\phi_{-}^2)-4c\beta\phi_{-}\nonumber\\
& &+ 2c^2+\xi^{-1}\big(1-\frac{1}{8}{c^2}-\frac{1}{128}c^4\big).
\label{eq:eq32}
\end{eqnarray}
From Eq.~(\ref{minimizec}) we can expand $c$ up to 3rd order in $\phi_{-}$:
$c\simeq a\phi_{-}+b\phi_{-}^3$.
Substituting this $c$ expression back into Eq.~(\ref{eq:eq32}) and retaining terms
up to fourth order in $\phi_{-}$, we can expand $\varepsilon(\phi_{+},\phi_{-})$
obtaining a fourth-order polynomial both in $\phi_{+}$ and $\phi_{-}$:
\begin{eqnarray}
\varepsilon(\phi_{+}, \phi_{-})&=&\phi_{-}^4\Big( 4ab-4b\beta-\frac{ab}{4}\xi^{-1}-\frac{a^4}{128}
\xi^{-1}+\frac{1}{2}\xi^{-2} \Big)\nonumber\\
&&+\phi_{-}^2\big(2a^2+2\beta^2-4a\beta-\frac{a^2}{8}\xi^{-1}\nonumber\\
&&+t\xi^{-2}+3\phi_{+}^2\xi^{-2} \big)\nonumber\\
&&+\frac{1}{2}\xi^{-2}\phi_{+}^4 +(2\beta^2+t\xi^{-2})\phi_{+}^2+\xi^{-1},
\label{eq:eq36}
\end{eqnarray}
where the coefficients $a$ and $b$ are defined as $a=16\beta\xi/(16\xi-1)$, and $b=a^3/(128\xi-8)$.

The free energy, Eq.~(\ref{eq:eq36}), is then minimized with respect to $\phi_{-}$, yielding
\begin{equation}
\label{eq:eq37}
\phi_{-}^2=(\delta-t\xi^{-2}-3\phi_{+}^2\xi^{-2})/\eta,
\end{equation}
with new coefficients $\delta$ and $\eta$ defined as $\delta=-2a^2-2\beta^2+4a\beta+ a^2\xi^{-1}/8 $
and $\eta=8ab-8b\beta- ab\xi^{-1}/2-a^4\xi^{-1}/64+\xi^{-2}$.
Substituting the expression of $\phi_{-}$ in Eq.~(\ref{eq:eq37}) into Eq.~(\ref{eq:eq36}), we get:
\begin{eqnarray}
\varepsilon(\phi_{+})&\simeq&-\frac{1}{2\eta}\Big(\delta -t\xi^{-2}- 3\xi^{-2}\phi_{+}^2 \Big)^2
+(2\beta^2+ t\xi^{-2})\phi_{+}^2  \nonumber\\
&&+\frac{1}{2}\xi^{-2}\phi_{+}^4+\xi^{-1}\, ,
\label{eq:eq38}
\end{eqnarray}
which is valid in the limit $c\ll 1$.

\begin{figure}[tbh]
\begin{center}
\includegraphics[scale=0.4]{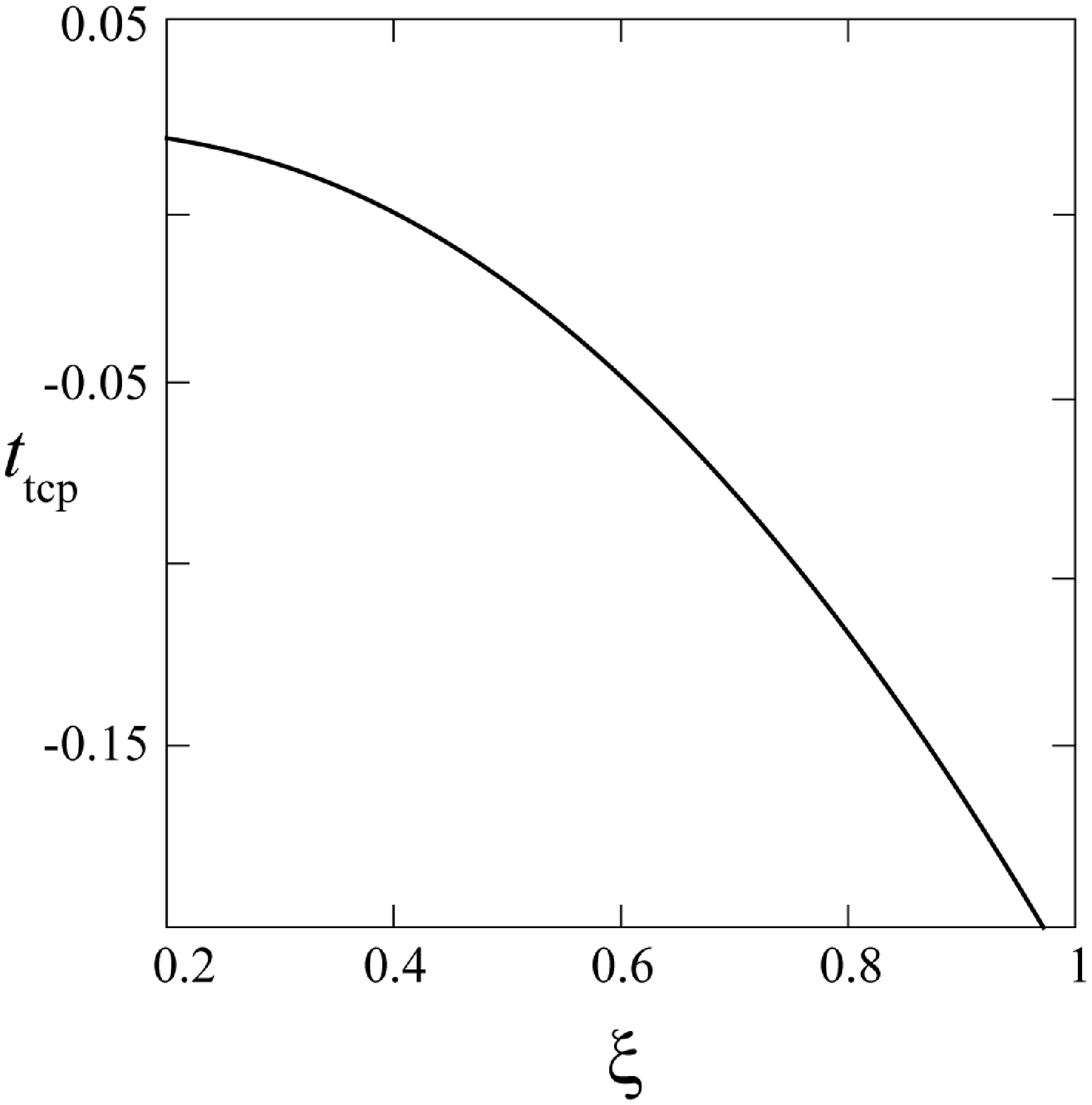}
\end{center}
\caption{\textsf{Tricritical temperature $t_{\rm tcp}$ as function of $\xi$,
calculated from Eq.~(\ref{eq:eq311}), for a fixed value of the coupling parameter, $\beta=1$.
The analytical expansion is valid only for $\xi$ values that are not too small.
In particular, below $\xi=0.2$ the overall change of the phase
diagram is from type I to II, as shown in Fig.~\ref{fig11}.
}}
\label{fig12}
\end{figure}

At the tricritical point, the free energies for $c=0$ and $c\ll 1$  are equal
for the same values of $\phi_{+}$ and $t$.
Comparing Eqs.~(\ref{eq:eq38}) and (\ref{eq:eq31}), we obtain the condition for the
tricritical point:
\begin{equation}
\delta - t\xi^{-2} - 3\xi^{-2}\phi_{+}^2 = 0.
\label{eq:eq39}
\end{equation}
In addition, the spinodal line is obtained from the requirement that
$\varepsilon^{\prime\prime}(\phi_{+})=0$ in Eq.~(\ref{eq:eq38}):
\begin{equation}
3\phi_{+}^2\Big(1-\frac{9}{\eta}\xi^{-2} \Big) + \frac{3}{\eta}\big(\delta - t\xi^{-2}\big)
+2\beta^2\xi^2+t=0.
\label{eq:eq310}
\end{equation}
By combining Eq.~(\ref{eq:eq39}) with Eq.~(\ref{eq:eq310}), we obtain:
\begin{equation}
t_{\rm{tcp}}=\xi^4(6\delta\xi^{-2}-\delta\eta-2\beta^2\eta)/6,
\label{eq:eq311}
\end{equation}
which depends on the values of $\xi$ and $\beta$.
In Fig.~\ref{fig12} we plot the variation of $t_{\rm tcp}$ with respect to $\xi$
for $\beta=1$ and $U/(2\kappa)=1$.
Substituting $\beta=1$ and $\xi=0.25$ in Eq.~(\ref{eq:eq311}),
we obtain $\phi_{+}^{\rm tcp}\simeq \pm 0.095$ and $t_{\rm tcp}\simeq 0.014$.
These tricritical values are in good agreement with the numerical ones, as can be read
off from the phase diagram of Fig.~\ref{fig5}(a),
$\phi_{+}^{\rm tcp}\simeq \pm 0.094$ and $t_{\rm tcp}\simeq 0.011$.

The present argument is valid only when $c \ll 1$ and can be applied to
type I phase diagram for which the domain size should be small (large $\xi$).
For type II phase diagram, on the other hand, the tricritical points are located
close to the fully-budded phase with $c=2$ (see Fig.~\ref{fig9}).
In such a case, the expansion in terms of $c$ cannot be justified,
and $t_{\rm tcp}(\xi)$ is no more valid for $\xi<0.2$ when $\beta=1$.


\begin{thebibliography}{99}

\bibitem{AlbertsBook}
B. Alberts, A. Johnson, P. Walter, J. Lewis, and M. Raff,
\textit{Molecular Biology of the Cell}
(Garland Science, New York, 2008).

\bibitem{VK05}
S. L. Veatch and S. L. Keller,
\textit{Biochim. Biophys. Acta} \textbf{1746}, 172 (2005).

\bibitem{SK_DA_Review}
S. Komura and D. Andelman,
\textit{Adv. Coll. Int. Sci.} \textbf{208}, 34 (2014).

\bibitem{VK}
S. L. Veatch and S. L. Keller,
\textit{Phys. Rev. Lett.} \textbf{94}, 148101 (2005).

\bibitem{YIMKO}
M. Yanagisawa, M. Imai, T. Masui, S. Komura, and T. Ohta,
\textit{Biophys. J.} \textbf{92}, 115 (2007).

\bibitem{UrsellBook}
T. S. Ursell,
\textit{Bilayer Elasticity in Protein and Lipid Organization}
(VDM Verlag, Berlin, 2009).

\bibitem{Lipowsky92}
R. Lipowsky,
\textit{J. Phys. II (France)} \textbf{2}, 1825 (1992).

\bibitem{Lipowsky93}
R. Lipowsky,
\textit{Biophys. J.} \textbf{64}, 1133 (1993).

\bibitem{hu}
J. Hu, T. Weikl, and R. Lipowsky,
\textit{Soft Matter} \textbf{7}, 6092 (2011).

\bibitem{JL93}
F. J\"ulicher and R. Lipowsky,
\textit{Phys. Rev. Lett.} \textbf{70}, 2964 (1993).

\bibitem{JL96}
F. J\"ulicher and R. Lipowsky,
\textit{Phys. Rev. E} \textbf{53}, 2670 (1996).

\bibitem{Lipowsky14}
R. Lipowsky,
\textit{Biol. Chem.} \textbf{395}, 253 (2014).

\bibitem{UKP}
T. S. Ursell, W. S. Klug, and R. Phillips,
\textit{Proc. Natl. Acad. Sci. U.S.A.} \textbf{106}, 13301 (2009).

\bibitem{baumgart}
T. Baumgart, S. T. Hess, and W. W. Webb,
\textit{Nature} \textbf{425}, 821 (2003).

\bibitem{RUPK}
J. E. Rim, T. S. Ursell, R. Phillips, and W. S. Klug,
\textit{Phys. Rev. Lett.} \textbf{106}, 057801 (2011).

\bibitem{HS}
W. Helfrich and R. M. Servuss,
\textit{Nuovo Cimento} \textbf{3}, 137 (1984).

\bibitem{lischi}
R. Lipowsky, M. Brickmann, R. Dimova, C. Haluska, J. Kierfeld, and J. Schillcock
\textit{J. Phys.: Condens. Matter} \textbf{17}, S2885 (2005).

\bibitem{SPA}
S. A. Safran, P. Pincus, and D. Andelman,
\textit{Science} \textbf{248}, 354 (1990).

\bibitem{SPAM}
S. A. Safran, P. A. Pincus, D. Andelman, and F. C. MacKintosh,
\textit{Phys. Rev. A} \textbf{43}, 1071 (1991).

\bibitem{MS}
F. C. MacKintosh and S. A. Safran,
\textit{Phys. Rev. E} \textbf{47}, 1180 (1993).

\bibitem{HM94}
J. L. Harden and F. C. MacKintosh,
\textit{Europhys. Lett.} \textbf{28}, 495 (1994).

\bibitem{harden}
J. L. Harden, F. C. MacKintosh, and P. D. Olmsted,
\textit{Phys. Rev. E} \textbf{72}, 011903 (2005).

\bibitem{GG01}
W. T. G\'o\'zd\'z and G. Gompper,
\textit{Europhys. Lett.} \textbf{55}, 587 (2001).

\bibitem{Ryu}
Y.-S. Ryu, I.-H. Lee, J.-H. Suh, S. C. Park, S. Oh,	
L. R. Jordan, N. J. Wittenberg,	S.-H. Oh, N. L. Jeon,	
B. Lee,	A. N. Parikh, and S.-D. Lee,
\textit{Nature Comm.} \textbf{5}, 4507 (2014).

\bibitem{Helfrich73}
W. Helfrich,
\textit{Z. Naturforsch. C} \textbf{28}, 693 (1973).

\bibitem{SafranBook}
S. A. Safran,
\textit{Statistical Thermodynamics of Surfaces, Interfaces, and Membranes},
(Addision Wesley, Reading, 1994).

\bibitem{SI}
K. Simons and E. Ikonen,
\textit{Nature} \textbf{387}, 569 (1997).

\bibitem{tian}
A. Tian, C. Johnson, W. Wang, and T. Baumgart,
\textit{Phys. Rev. Lett.} \textbf{98}, 208102 (2007).

\bibitem{BEG}
M. Blume, V. J. Emery, and R. B. Griffiths,
\textit{Phys. Rev. A} \textbf{4}, 1071 (1971).

\bibitem{LS}
D. Lingwood and K. Simons,
\textit{Science} \textbf{327}, 46 (2010).

\bibitem{SA}
M. Seul and D. Andelman,
\textit{Science} \textbf{267}, 476 (1995).

\bibitem{leibler}
S. Leibler and D. Andelman,
\textit{J. Physique (Paris)} \textbf{48}, 2013 (1987).

\bibitem{andel}
D. Andelman, T. Kawakatsu, and K. Kawasaki
\textit{Europhys. Lett.} \textbf{19}, 57 (1992).

\bibitem{taniguchi}
T. Taniguchi, K. Kawasaki, D. Andelman, and T. Kawakatsu,
\textit{J. Phys. II (France)} \textbf{4}, 1333 (1994).

\bibitem{kawa}
T. Kawakatsu, D. Andelman, K. Kawasaki, and T. Taniguchi,
\textit{J. Phys. II (France)} \textbf{3}, 971 (1993).

\bibitem{KK}
H. Kodama and S. Komura,
\textit{J. Phys. II (France)} \textbf{3}, 1305 (1993).

\bibitem{kumar}
P. B. Sunil Kumar, G. Gompper, and R. Lipowsky,
\textit{Phy. Rev. E} \textbf{60}, 4610 (1999).

\bibitem{BPS09}
R. Brewster, P. A. Pincus, and S. A. Safran,
\textit{Biophys. J.} \textbf{97}, 1087 (2009).

\bibitem{YBS}
T. Yamamoto, R. Brewster, and S. A. Safran,
\textit{EPL} \textbf{91}, 28002 (2010).

\bibitem{PS13}
B. Palmieri and S. A. Safran,
\textit{Langmuir} \textbf{29}, 5246 (2013).

\bibitem{PS14}
B. Palmieri and S. A. Safran,
\textit{Langmuir} \textbf{30}, 11734 (2014).

\bibitem{hirose09}
Y. Hirose, S. Komura, and D. Andelman,
\textit{Chem. Phys. Chem.} \textbf{10}, 2839 (2009).

\bibitem{hirose12}
Y. Hirose, S. Komura, and D. Andelman,
\textit{Phys. Rev. E} \textbf{86}, 021916 (2012).

\bibitem{shlomovitz}
R. Shlomovitz, L. Maibaum, and M. Schick,
\textit{Biophys. J} \textbf{106}, 1979 (2014).

\bibitem{SH}
A. Shioi and T. A. Hatton,
\textit{Langmuir} \textbf{18}, 7341 (2002).

\end{thebibliography}

\end{document}